\documentclass[11pt,twoside]{article}
\usepackage{FUSE2004}
\usepackage{natbib}

\usepackage{epsf}
\usepackage{psfig}
\usepackage{lscape}

\begin{document}
\title{AA\,Dor -- An Eclipsing Post Common-Envelope Binary}
\author{T. Rauch}
\affil{Dr.-Remeis-Sternwarte, 96049 Bamberg, Germany\\
       Institut f\"ur Astronomie und Astrophysik, 72076 T\"ubingen, Germany}
\author{K. Werner}
\affil{Institut f\"ur Astronomie und Astrophysik, 72076 T\"ubingen, Germany}

\begin{abstract}
AA\,Dor (LB\,3459) is an eclipsing, close, single-lined, 
post common-envelope binary (PCEB)
consisting of an sdOB primary star and an unseen secondary with an 
extraordinary small mass -- formally a brown dwarf. The brown dwarf 
may have been a former planet which survived a common envelope 
phase and has even gained mass.

A recent determination of the components' masses from results of 
state-of-the-art NLTE spectral analysis and subsequent comparison 
to evolutionary tracks shows a discrepancy between masses derived from 
radial-velocity and the eclipse curves. Phase-resolved high-resolution 
and high-SN spectroscopy was carried out with FUSE in order to 
investigate on this problem. 

We present preliminary results of an ongoing NLTE spectral analysis of FUSE spectra of the 
primary.
\end{abstract}

\section{Introduction}
A recent spectral analysis of the primary of AA\,Dor (Rauch 2000, 2004) has
shown a discrepancy between the resulting stellar parameters from spectral 
analysis and those derived from the radial-velocity and the eclipse curves. 
Recently, Hilditch et al\@. (2003) presented new photometry data and an 
improved photometric model of AA\,Dor which
verified earlier results within smaller error ranges.
Thus, one reason for this discrepancy might be uncertainties in the spectral analysis, i.e\@.
mainly in the surface gravity determination. 

Phase-resolved high-resolution UV spectra are
necessary in order to reduce the error limits and to identify individual iron and 
nickel lines in the spectrum. 
Ten LWRS observations with short exposure times (200\,s) should be performed,
in order to minimize the effects of orbital smearing.
Four have already been executed on Aug 29, 2003. Five 
were carried out on Jun 22, 2004. The preliminary analysis which is presented
here makes use of the 2003 data, only.

\begin{figure}[ht]
\plotone{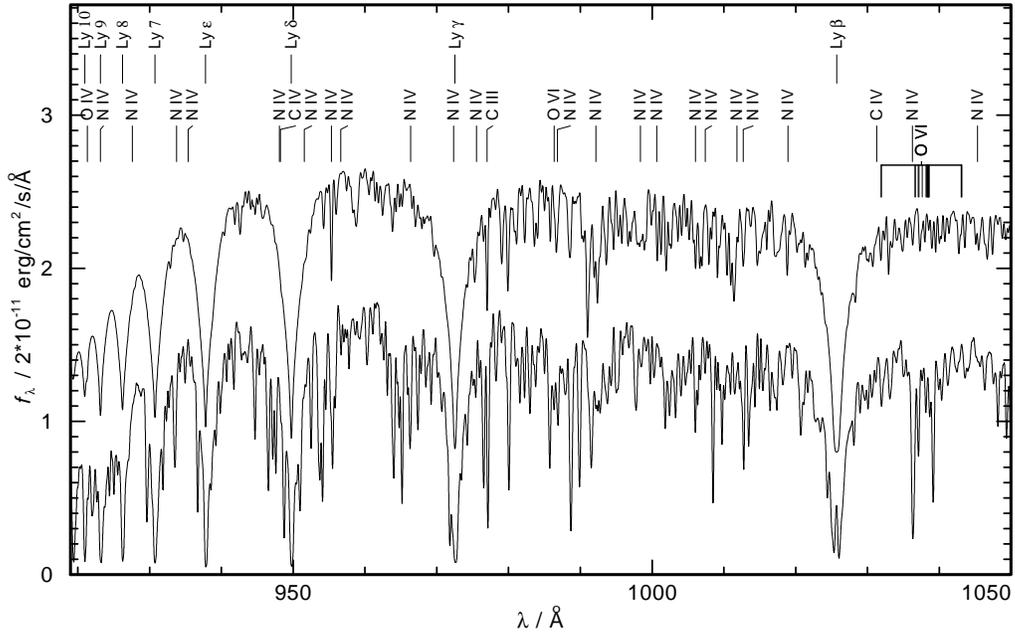}
\caption{Comparison of the four co-added FUSE spectra from Aug 2003 (bottom)
of the sdOB primary of AA\,Dor with the synthetic spectrum
of our final model (top, Rauch 2000). Both are smoothed with a Gaussian of 0.2\,\AA\ FWHM for clarity. 
H, C, N, and O lines considered in the model are indicated. 
All measured Kurucz lines (POS, Kurucz 1996) are considered.
The synthetic spectrum is convolved with a rotational profile with 35 km/sec. 
}
\end{figure}

The plane-parallel, static models (H+He+C+N+O+Mg+Si+IronGroup included), 
used for the spectral analysis of AA\,Dor, 
are calculated with TMAP, the T\"ubingen NLTE Model Atmosphere Package
(Werner et al\@. 2003, Rauch \& Deetjen 2003).

\begin{figure}[htb]
\plotone{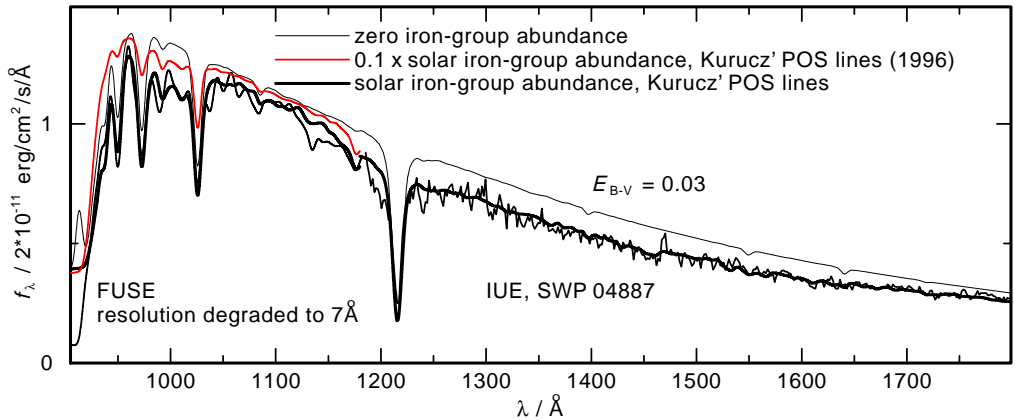}
\caption{Total abundance of the iron-group elements.
We reduced the resolution of the FUSE spectrum to 7\,\AA\ (like the low-resolution
IUE spectrum SWP\,04887 for wavelength greater than 1186\,\AA).
The overall slope of the flux is matched well
by a solar abundance of the iron-group elements.
}
\end{figure}

\clearpage

\begin{figure}[ht]
\plotone{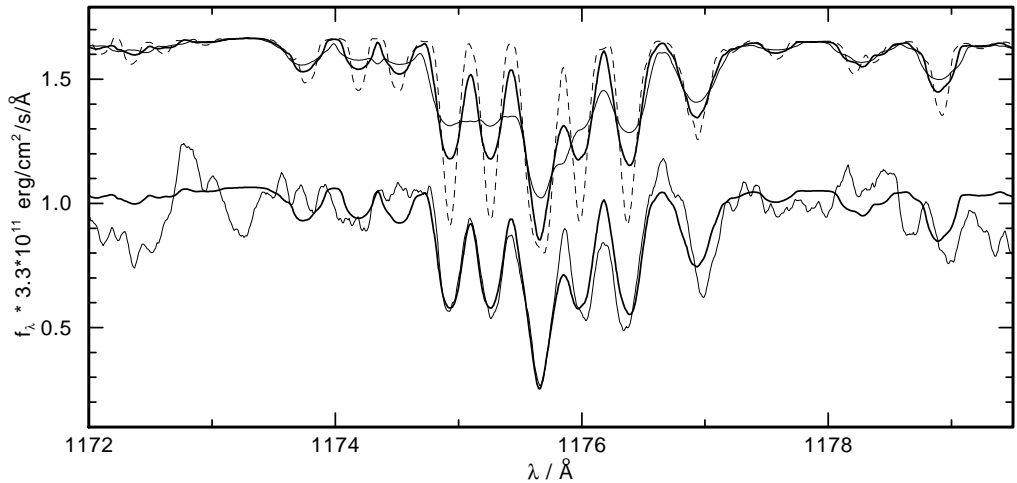}
\caption{Rotational velocity of the primary AA\,Dor.
The FUSE spectrum around C\,{\sc iii} 1175\,\AA\ is compared with a synthetic spectrum which is
convolved with a rotational profile with $v_\mathrm{rot}=35\,\mathrm{km/sec}$ (bottom). 
The top panel shows three synthetic spectra calculated with $v_\mathrm{rot}=20, 35, 50\,\mathrm{km/sec}$
(dashed, thick, thin line).}
\end{figure}

\section{Results}

From the C\,{\sc iii} 1175\,\AA\ triplet, we have determined a rotational velocity of
35 +/- 10\,km/s for the primary of AA\,Dor (Fig.\,3). This is lower than 
47 +/- 5\,km/sec found by Rauch \& Werner (2003).
The total abundance of the iron-group elements (Fig.\,2), derived from the overall
flux shape is about solar. The same result was found by Rauch (2000).
However, we are not able to identify any iron-group line unambiguously (Fig.\,1).

\acknowledgements{This research was supported by the DLR under grant 50\,OR\,0201.}

\end{document}